\documentclass[aps,prb,showpacs,twocolumn,citeautoscript,superscriptaddress,footinbib,bibnotes]{revtex4-2}
\usepackage[dvips]{graphicx}
\usepackage{dcolumn}
\usepackage{bm}
\usepackage{fancyhdr}
\usepackage{placeins}
\usepackage{ifthen}
\usepackage{eurosym}
\usepackage{float}
\usepackage{longtable}
\usepackage{hyperref}
\usepackage{footnote}
\usepackage[english]{babel}
\usepackage[usenames,dvipsnames]{color}

\begin{document}


\title{YBa$_{1-x}$Sr$_{x}$CuFeO$_{5}$ layered perovskites: exploring the magnetic order beyond the paramagnetic-collinear-spiral triple point}
\vspace{-7mm}

\author{V.~Por\'ee}
\email{victor.poree@psi.ch}
\affiliation{Laboratory for Multiscale Materials Experiments, Paul Scherrer Institut, 5232 Villigen PSI, Switzerland}
\affiliation{Synchrotron SOLEIL, L'Orme des Merisiers, Saint-Aubin, BP 48, F-91192 Gif-sur-Yvette, France}
\author{D.J.~Gawryluk}
\email{dariusz.gawryluk@psi.ch}
\affiliation{Laboratory for Multiscale Materials Experiments, Paul Scherrer Institut, Villigen 5232 Villigen PSI, Switzerland}

\author{T.~Shang}
\affiliation{Laboratory for Multiscale Materials Experiments, Paul Scherrer Institut, 5232 Villigen PSI, Switzerland}
\affiliation{Key Laboratory of Polar Materials and Devices (MOE), School of Physics and Electronic Science, East China Normal University,Shanghai 200241, China}

\author{J.A.~Rodr\'iguez-Velamaza\'n}
\affiliation{Institut Laue-Langevin, 6 rue Jules Horowitz, BP 156, 38042 Grenoble Cedex 9, France}

\author{N.~Casati}
\affiliation{Swiss Light Source, Paul Scherrer Institut, 5232 Villigen PSI, Switzerland}

\author{D.~Sheptyakov}
\affiliation{Laboratory for Neutron Scattering and Imaging, Paul Scherrer Institut, 5232 Villigen PSI, Switzerland}

\author{X.~Torrelles}
\affiliation{ICMAB CSIC, Institut de Ci\`{e}ncia de Materials de Barcelona, Campus UAB, Bellaterra 08193, Catalonia, Spain}

\author{M.~Medarde}
\email{marisa.medarde@psi.ch}
\affiliation{Laboratory for Multiscale Materials Experiments, Paul Scherrer Institut, 5232 Villigen PSI, Switzerland}

\date{\today}

\begin{abstract}
Layered perovskites of general formula AA'CuFeO$_5$ are one of the few examples of cycloidal spiral magnets where the ordering temperatures $T_{spiral}$ can be tuned far beyond room temperature by introducing modest amounts of Cu/Fe chemical disorder in the crystal structure. This rare property makes these materials prominent candidates to host multiferroicity and magnetoelectric coupling at room temperature. Moreover, it has been proposed that the highest $T_{spiral}$ value that can be reached in this structural family ($\sim$~400~K) corresponds to a paramagnetic-collinear-spiral triple point with potential to show exotic physics. Since generating high amounts of Cu/Fe disorder is experimentally difficult, the phase diagram region beyond the triple point has been barely explored. To fill this gap we investigate here the YBa$_{1-x}$Sr$_{x}$CuFeO$_{5}$ solid solutions ($0~\leq~x~\leq~1$), where we replace Ba with Sr with the aim of enhancing the impact of the experimentally available Cu/Fe disorder. Using a combination of bulk magnetization, synchrotron X-ray and neutron powder diffraction we show that the spiral state is destabilized beyond a critical degree of Cu/Fe disorder, being replaced by a non-frustrated, fully antiferromagnetic state with propagation vector \textbf{k$_{c2}$}~=~$(\frac{1}{2},~\frac{1}{2},~0)$ and ordering temperature $T_{coll2}$~$\geq$~$T_{spiral}$, which is progressively stabilized beyond the triple point. Interestingly, $T_{spiral}$ and $T_{coll2}$ increase with $x$ at the same rate. This suggests a common, disorder-driven origin, consistent with theoretical predictions. 
\end{abstract}

\maketitle


\section{Introduction}

Magnetoelectric multiferroics, materials with coupled magnetic and ferroelectric orders~\cite{curie,Fiebig,Spaldin1}, have attracted significant attention during the last years owing to their potential for applications in spintronics and low-power data storage technologies~\cite{Multifer_app}. Among them, frustrated magnets with ordered magnetic spiral phases that spontaneously break inversion symmetry~\cite{Goto1,Katsura1,Cheong1} have received particular attention due to their capability to induce ferroelectricity~\cite{Heyer_MnWO4,Kimura_2003,PRL_Ferro_spiralMagent,Ajay_FeVO4}. In these materials, the common origin of both orders guarantees a substantial coupling between them, which is highly desirable for applications. However, spiral magnetic orders are often the signature of competing magnetic interactions, and in this case, their ordering temperatures $T_{spiral}$ are limited by the energy scale of the weakest magnetic coupling. With very few exceptions~\cite{Kimura_CuO,Giovannetti_CuO}, this results in low $T_{spiral}$ values (typically $<$~100~K), greatly restricting their fields of application.

\begin{figure}[H]
	\centering
    \includegraphics[width=0.50\textwidth]{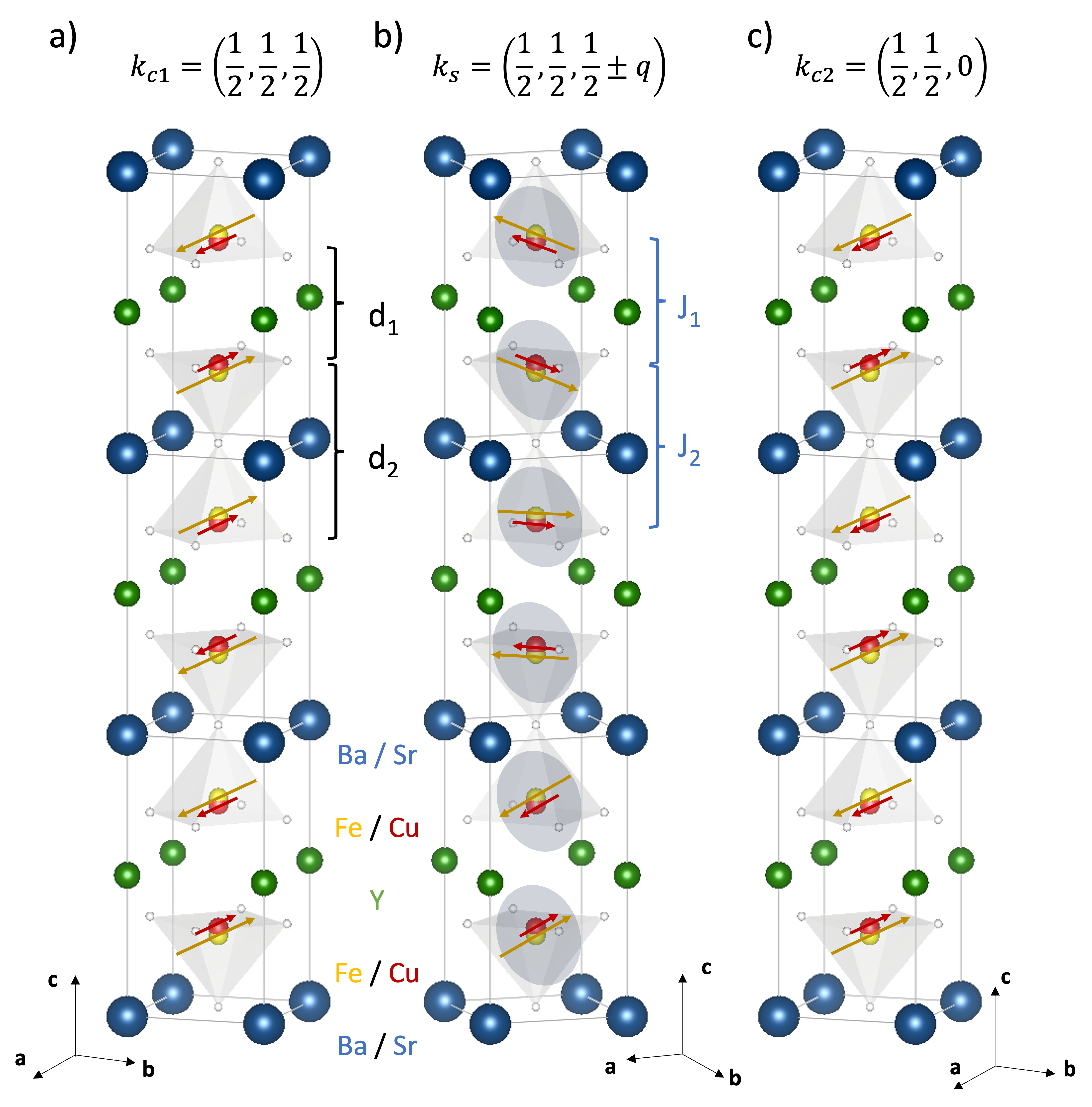}
	\caption{The three magnetic structures observed in the YBa$_{1-x}$Sr$_x$CuFeO$_5$ series. a) Collinear with \textbf{k$_{c1}$} = $(\frac{1}{2}, \frac{1}{2}, \frac{1}{2})$. b) Inclined circular spiral with \textbf{k$_{s}$} = $(\frac{1}{2},  \frac{1}{2},  \frac{1}{2}\pm\textit{q})$. c) Collinear structure with propagation vector \textbf{k$_{c2}$} = $(\frac{1}{2}, \frac{1}{2}, 0)$. Oxygen atoms are depicted as little white spheres, and grey lines delimit the crystallographic unit cell. For clarity, only one crystal cell within the \textbf{ab} plane is shown. Red and yellow arrows denote Cu$^{2+}$ and Fe$^{3+}$ magnetic moments, respectively. The interatomic distances $d_1$, $d_1$ and magnetic couplings $J_1$, $J_1$ mentioned in the text are also indicated.}
	\label{Structure}
\end{figure}

Recently, investigations on Cu/Fe-based layered perovskites uncovered an unexpected knob to control the stability range  of a magnetic spiral: chemical disorder. The positive impact of this variable in the spiral ordering temperature was first demonstrated for YBaCuFeO$_5$~\cite{Morin2}, a simple tetragonal material whose crystal structure is shown in Fig.~\ref{Structure}~\cite{Rakho,Morin1}. This layered perovskite contains two pseudocubic ABO$_3$ units due to the ordering of the A-cations Y$^{3+}$ and Ba$^{2+}$ along the \textbf{c} crystal axis. This results in a tetragonal unit cell $a_c$~x~$a_c$~x~$c$ ($c$~$\simeq$~2$a_c$), where $a_c$ denotes the pseudocubic perovskite lattice parameter. Cu$^{2+}$ and Fe$^{3+}$ occupy the B-positions, which form layers of apex-linked bipyramids due to the existence of 1/6 of oxygen vacancies in the Y layers. Contrarily to the A-cations, Cu$^{2+}$ and Fe$^{3+}$ present important amounts of preparation-dependent disorder~\cite{Morin2,Morin1,TianScienceM,Zhang_2021,Romaguera_2022,Zhang_2022, Marelli_2024}. This is evidenced by the variable Cu/Fe occupations $n_{Cu}$ and $n_{Fe}$ of the square-pyramidal sites reported in diffraction studies, which represent respectively the probabilities of having a given pyramid occupied by Cu$^{2+}$ and Fe$^{3+}$ ($n_{Cu}$~=~$n_{Fe}$~=~50$\%$ in the case of maximal intermixing). This chemical disorder is illustrated in Fig.~\ref{Structure} by the splitting of the B-cation position with the pyramids, where the slightly different Cu and Fe $z$-coordinates also reflect the Jahn-Teller nature of Cu$^{2+}$ ($t^6_{2g}$~$e^3_g$) and its preference for long Cu-O apical distances ~\cite{Capponi_1987,Meyer_1990,Huang_1992,Morin1, Marelli_2024}. 

YBaCuFeO$_5$ displays long-range antiferromagnetic (AFM) order up to elevated temperatures, and undergoes two phase transitions ~\cite{Caignaert_1995,Morin1}. At high temperature the magnetic structure is collinear (henceforth $coll1$, Fig.~\ref{Structure}a), but it transforms into an inclined spiral at lower temperatures (Fig.~\ref{Structure}b). Remarkably, the setup of the spiral phase can be increased from 154~K to 400~K upon a modest enhancement in the degree of Cu/Fe intermixing~\cite{Morin1,Morin2,Romaguera_2022}. Since this parameter has the opposite impact on the collinear ordering temperature, the two transitions eventually meet by continuously increasing the degree of Cu/Fe intermixing. According to previous studies this happens at $T_{coll1}$~=~$T_{spiral}$~$\sim$~400~K~\cite{TianScienceM,Romaguera_2022}, giving rise to a paramagnetic-collinear-spiral triple point. These studies also suggest that the spiral is destabilized by further increasing the degree of Cu/Fe disorder, in which case the triple point would define the highest $T_{spiral}$ value that can be achieved in YBaCuFeO$_5$ \cite{TianScienceM,Romaguera_2022}. Giving the scarcity of materials with magnetic spirals stable far beyond room temperature (RT), clarifying this issue is highly desirable. However, generating the disorder amounts needed to access the region beyond the triple point has been technically challenging and so far unsuccessful.

Here, we investigate this region using an alternative route aimed to enhance the impact of the maximal disorder accessible to our experimental setup. As shown in our previous study~\cite{TianScienceM}, the degree of magnetic frustration - and hence $T_{spiral}$ - can be also enhanced by increasing the magnetic coupling $J_2$ between the B-cations occupying the bipyramidal units of the crystal structure (Fig.~\ref{Structure}b), an effect that adds up to that of disorder. To reach this goal we partially replace Ba$^{2+}$ with Sr$^{2+}$, an isovalent cation with smaller ionic radius that reduces the size $d_2$ of the bipyramids (Fig.~\ref{Structure}a), effectively increasing $J_2$.  Using an improved synthetic route aimed to create samples with similar Fe/Cu disorder and increasingly growing $J_2$ values we successfully prepared the full YBa$_{1-x}$Sr$_x$CuFeO$_5$ solid solution (0~$\leq$~$x$~$\leq$~1), extending the range of our previous study (0~$\leq$~$x$~$\leq$~0.5)~\cite{TianScienceM} and earlier publications~\cite{Yadav1,Yadav2,Yadav3}. We use here a combination of bulk SQUID magnetometry, neutron and synchrotron X-ray powder diffraction, that we employ to investigate the evolution of the crystal structures and the stability of the different magnetic phases. Our analysis reveals the existence of a maximal $T_{spiral}$ value close to 260~K, which corresponds to $x_{t}$~$\sim$~0.5. It also uncovers the disappearance of the spiral phase for $x$~$>$~$x_{t}$ and its replacement by a new, distinct collinear AFM phase (henceforth $coll2$), shown in Fig.~\ref{Structure}c. Intriguingly, the ordering temperature $T_{coll2}$ of the new phase increases with $x$ at a rate $\partial$$T_{coll2}$/$\partial$$x$ nearly identical to that of $T_{spiral}$ in the $x~<~x_{t}$ region. Moreover, it crosses the $T_{coll1}$ line when $T_{coll1}$~=~$T_{coll2}$~$\sim$~400~K. These observations suggest a common origin for the spiral and $coll2$ phases, that we discuss within the framework of the random exchanges model recently proposed by Scaramucci and co-workers~\cite{PRX.8,scaramucci2016spiral}.

\section{Experimental Methods}

\subsection{Material synthesis}

The eleven YBa$_{1-x}$Sr$_{x}$CuFeO$_{5}$ polycrystalline samples used in this study (0~$\leq$~$x$~$\leq$~1 in steps of 0.1) were prepared by solid state reaction. Some of these compositions have been already synthesized in the past ~\cite{TianScienceM,Yadav1,Yadav2,Yadav3}, but the conditions employed in previous studies did not allow to stabilize materials with high Sr contents ($x$~$\geq$~0.5) as a single phase. Here we follow an improved, two-step methodology where the synthesis is preceded by a thermal stability investigation of each individual composition using Simultaneous Thermal Analysis (STA) by means of Thermogravimetry and Differential Thermal Analysis (TGA-DTA). This technique allows to explore the impact of the heating rate, annealing atmosphere, thermal stability regions and presence of eventual phase transitions during the solid state synthesis process. Moreover, when combined with powder X-ray diffraction, it enables a fast, precise determination of the minimal temperature needed for a material's formation, as well as its thermal stability range upon heating.

As precursors we employed high-purity Y$_{2}$O$_{3}$ (99.99\%, (REO)-STREM CHEMICALS, previously pre-annealed at 950$^{\circ}$C for 10h), BaCO$_{3}$(Puratronic, 99.997\%, Metal basis, Alfa Aesar), SrCO$_{3}$(Puratronic, 99.994\%, Metal basis, Alfa Aesar), CuO$_{2}$(Puratronic, 99.995\%, Metal basis, Alfa Aesar) and Fe$_{2}$O$_{3}$(Puratronic, 99.998\%, Metal basis, Alfa Aesar). For each compound, about 10~g of precursors were weighted and mixed in stoichiometric amounts. Small portions of the precursor mixtures ($\simeq$~200~mg) were first loaded in alumina holders and subject to thermal analysis (TA) measurements on a NETZSCH 449C Simultaneous Thermal Analyzer (STA). Mass loss and DTA curves were subsequently measured between room temperature (RT) and 1400$^{\circ}$C in artificial air gas flow using a heating rate of 2.5$^{\circ}$C/min. Additional curves were measured with isothermal steps below the melting points in order to investigate the nature of the stable crystalline phases. The phase composition of the obtained products was assessed from the analysis of the X-ray powder diffraction (XRPD) patterns measured using a Bruker D8 Advance powder diffractometer (Cu $K_{\alpha}$).

\begin{figure}[h!]
	\centering
    \includegraphics[width=0.48\textwidth]{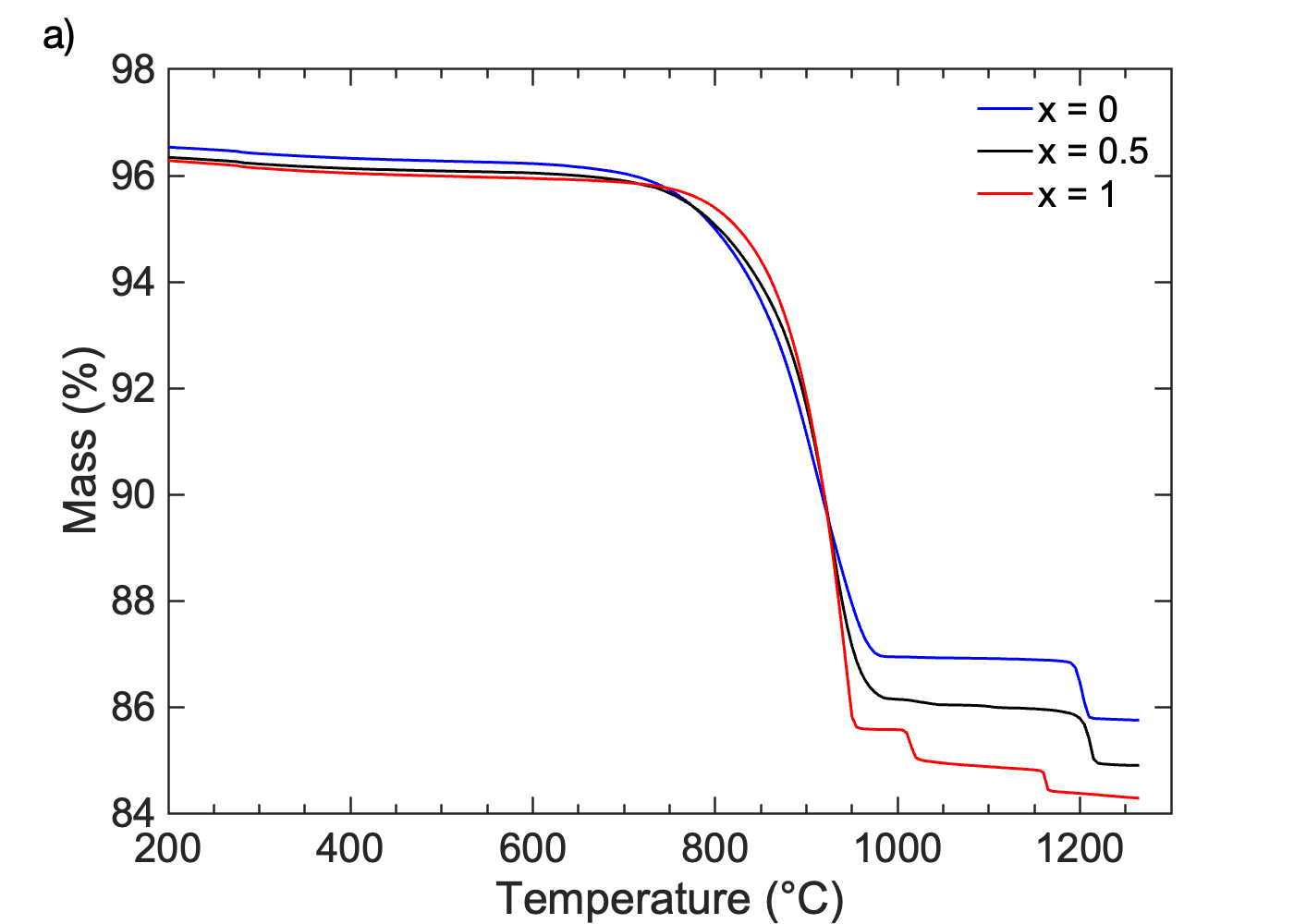}	\includegraphics[width=0.48\textwidth]{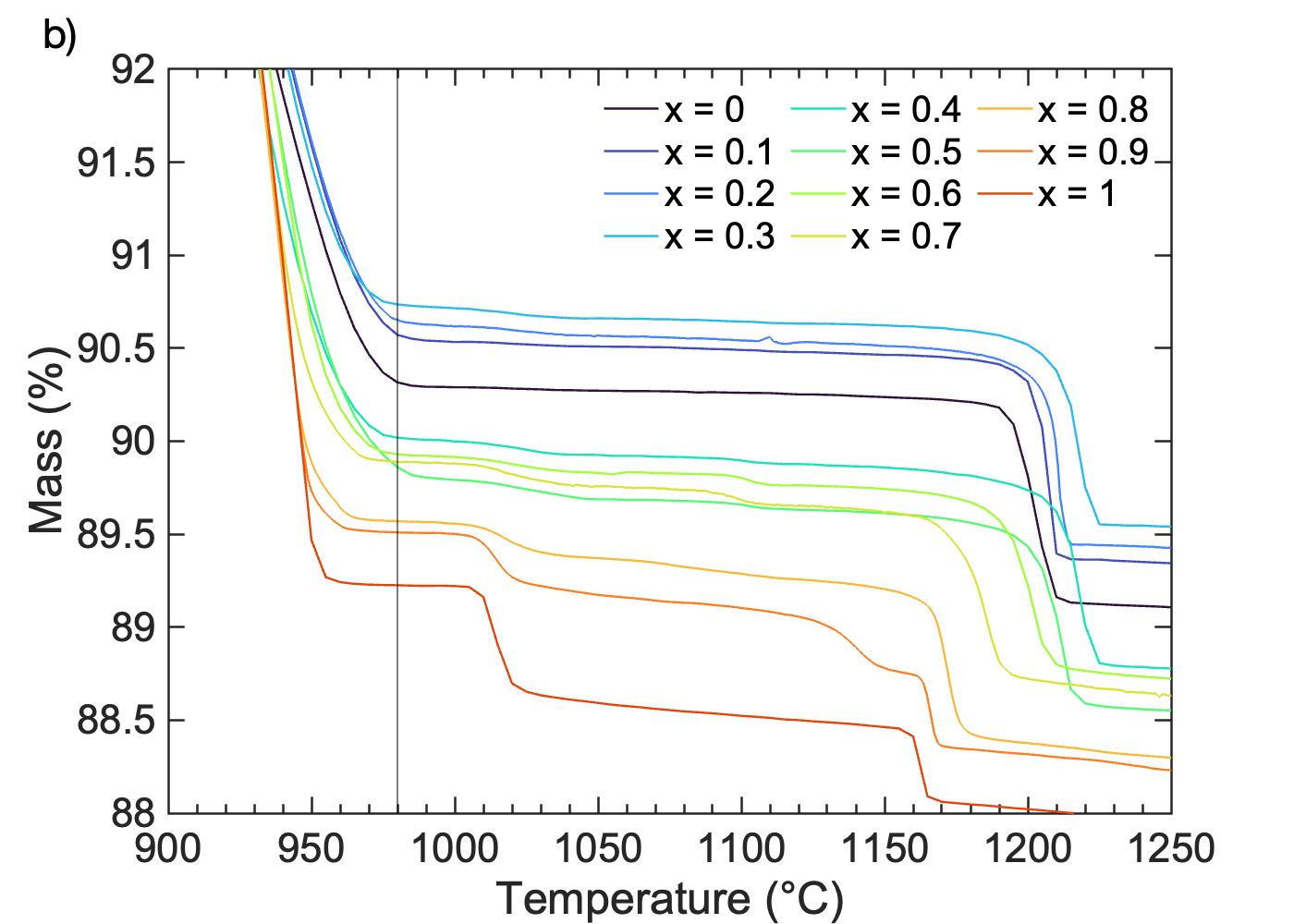}
    \includegraphics[width=0.48\textwidth]{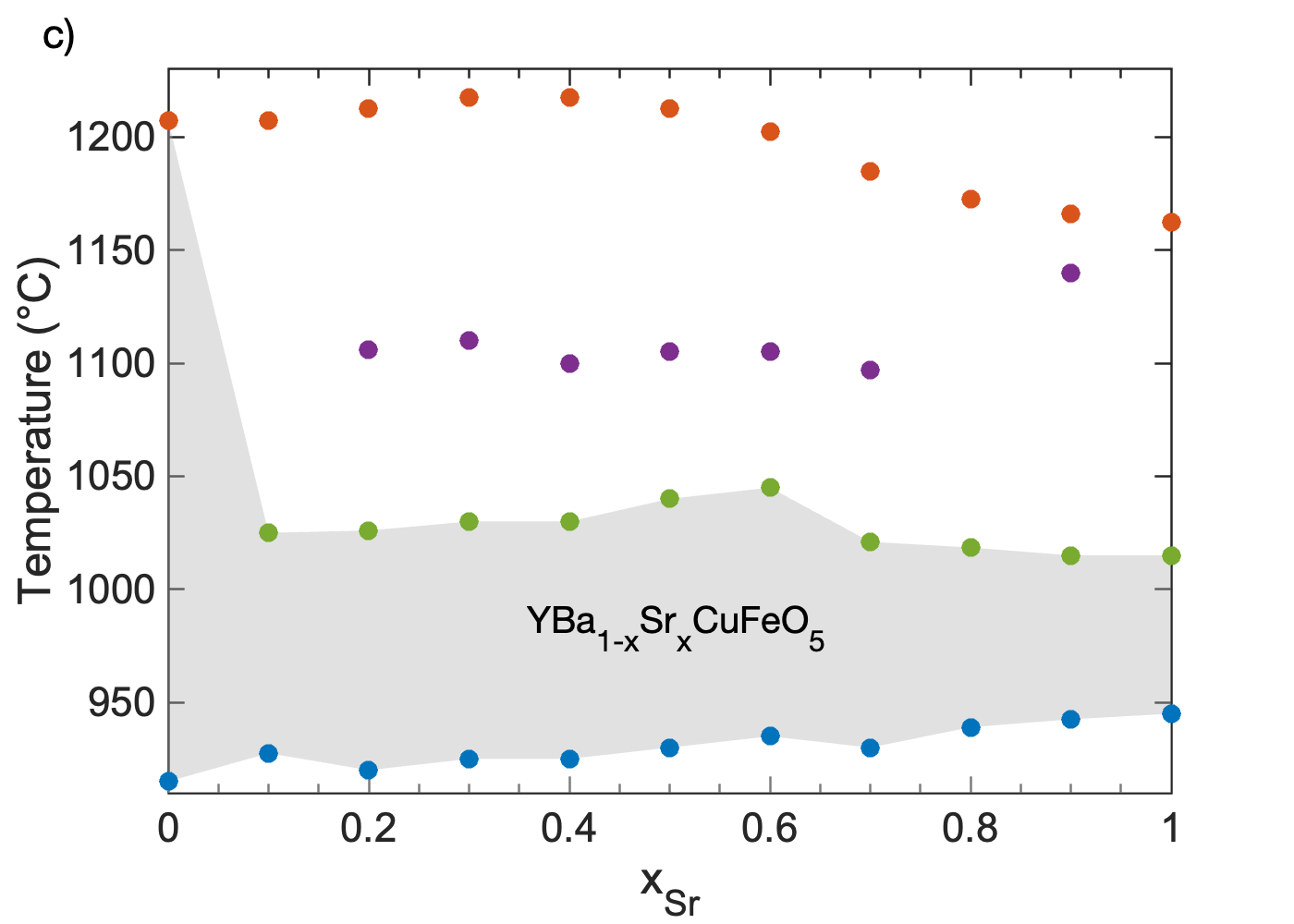}
	\caption{a) TG curves of the YBa$_1-x$Sr$_x$CuFeO$_5$ samples with $x$~=~0, 0.5, 1. b) Detail of the plateau region for the eleven compositions investigated in this study. The black vertical line is a guide for the eye, highlighting the chosen synthesis temperature. c) Evolution of the layered perovskite stability region for increasing Sr content. The blue and orange markers indicate respectively the beginning of the phase formation and the phase melting. Green and purple markers indicate the formation of other, competing crystalline phases. All points correspond to inflection points in the TG curves.}
	\label{TGs}
\end{figure}

Fig.~\ref{TGs}a shows three representative TG curves corresponding to the starting, middle and end members of the family YBaCuFeO$_5$ ($x$~=~0), YBa$_{0.5}$Sr$_{0.5}$CuFeO$_5$ ($x$~=~0.5) and YSrCuFeO$_5$ ($x$~=~1). The curves reveal a continuous mass loss upon heating followed by one or multiple plateaus, whose fine structure is shown in more detail in Fig.~\ref{TGs}b together with the curves of the remaining samples. For the Sr-free material YBaCuFeO$_5$ we observe a single, broad plateau starting at $T_{start}$~$\simeq$~920$^{\circ}$C and ending at $T_{end}$~$\simeq$~1200$^{\circ}$C that we define here as the temperatures where an inflection point is found in the derivative of the TG curves. According to XRPD, $T_{start}$ and $T_{end}$ correspond to new phases formation accompanied by their respective mass loss. They delimit thus the stability region of the layered perovskite structure (Fig.~\ref{TGs}c). Upon introduction of strontium in the structure, steps of increasing height appear in the original plateau, suggesting the formation of other stable phases (Supplemental Material). This was confirmed by XRPD, which revealed a strong decrease of the layered perovskite stability region by increasing the Sr content (from $\simeq$~280~degrees to $\simeq$~60~degrees, see Fig.~\ref{TGs}c).

According to previous studies~\cite{Caignaert_1995,Morin1,Morin2,TianScienceM,Zhang_2021,Zhang_2022,Romaguera_2022, Marelli_2024}, the degree of Cu/Fe disorder in the structure is controlled by two main variables: the annealing temperature, and the cooling rate of the last annealing. To obtain large amounts of disorder (and hence $T_{spiral}$ values), these two variables should thus be as high as possible. In the case of the YBa$_{1-x}$Sr$_{x}$CuFeO$_{5}$ solid solutions, where we were aiming to introduce similar Cu/Fe disorder amounts, the reduced stability range of the Sr-rich compounds limits considerably the choice of the annealing temperature. After some preliminary tests we chose an annealing temperature of 980$^{\circ}$C. This value was substantially lower than that employed in our previous study (1150$^{\circ}$C, ~\cite{TianScienceM}), but was a good compromise allowing to minimize the amount of impurities while still producing a fair degree of Cu/Fe disorder.

For the elaboration of the final materials, the weighted and mixed precursors were pressed into pellets, sintered in air at 980$^{\circ}$C for 48h, and cooled down in the oven to RT. The mixing, pelletizing and sintering process was repeated seven times, with XRPD patterns recorded after every cycle in order to keep track of the samples purity through the cyclic process. After the last annealing all samples were quenched into liquid nitrogen in order to maximize their Cu/Fe disorder. Small (50-60~mg) pieces of the final pellets were kept solid for their use in magnetization measurements, and the rest were pulverized for neutron and synchrotron X-ray powder diffraction experiments. XRPD analysis indicated that all the obtained materials were very well crystallized, and most of them (0~$<$~$x$~$\leq$~0.9) contained only minor amounts of Y$_2$Cu$_2$O$_5$. In the case of the end member of the series ($x$~=~1), where the impurity content is more important, we observe also the signature of Y$_2$SrCuFeO$_{6.5}$ and CuO (Supplemental Material).

\subsection{Magnetometry}
The presence of magnetic order in the YBa$_{1-x}$Sr$_x$CuFeO$_5$ samples was first assessed using a superconducting quantum interference magnetometer equipped with oven (MPMS XL, Quantum Design). DC magnetization ($M$) measurements as a function of temperature were carried out under an external magnetic field of $H$~=~0.5~T by heating after cooling the samples in zero-field. For the low temperature range (10~K~$\leq$~$T$~$\leq$~400~K), the samples were mounted in transparent drinking straws, whose contribution to the total magnetization was negligible. For measurements at higher temperatures (300~K~$\leq$~$T$~$\leq$~500~K) they were wrapped in Al foil and mounted as described in ref. ~\cite{Sese_2007}. The contribution of the Al and the oven insert was measured separately in the same conditions and subtracted from the data. The magnetic susceptibility $\chi~=~M/H$ was subsequently calculated for all samples.

\subsection{Neutron and Synchrotron X-ray diffraction}
The nature and thermal stability of the different magnetically ordered phases was further investigated using neutron powder diffraction (NPD). All measurements were performed on the high-intensity powder diffractometer D1B located at the Institut Laue-Langevin, Grenoble (France)~\cite{Diffracto}. The powderized samples were introduced in vanadium cans ($D$~=~0.6~cm, $H$~=~5~cm), mounted in a cryofurnace, and cooled down to 2~K. NPD patterns were then recorded up to about 500~K by heating at a constant rate of 1.5~K/min (Pyrolytic Graphite, PG (002), $\lambda$~=~2.52~\AA, 2$\theta_{max}$~=~128$^{\circ}$, 2$\theta_{step}$~=~0.1$^{\circ}$). The acquisition time of a single pattern was typically 3 minutes, and the background from the cryofurnace was minimized using an oscillating radial collimator. High statistics patterns were measured at 2~K and 300~K, and employed for detailed magnetic structure refinements. Additional high statistics patterns were recorded at the same temperatures using a shorter wavelength (Ge (311), $\lambda$~=~1.28~\AA). The wavelengths and zero offsets were determined using a Na$_2$Ca$_3$Al$_2$F$_{14}$ (NAC) reference powder sample.

The evolution on the crystal structure was further investigated using Synchrotron X-ray powder diffraction (SXPD). The measurements were conducted at the Swiss Light Source (SLS) of the Paul Scherrer Institute in Villigen, Switzerland~\cite{MS}. Room temperature diffraction patterns of all samples were measured on the Materials Science beamline ($\lambda$~=~0.4959369~\AA, 2$\theta_{max}$~=~75$^{\circ}$, 2$\theta_{step}$~=~0.0036$^{\circ}$). The powder samples were finely ground, loaded in borosilicate glass capillaries ($D$~=~0.1~mm,  0.32~$\leq$~$\mu$$R$~$\leq$~0.37) and measured in transmission mode with a rotation speed of $\sim$~2~Hz using a Mythen II 1D multistrip detector (Dectris). The primary beam was vertically focused and slitted to about 300~x~4'000~$\mu$m$^2$. The wavelength and zero offset were determined using a LaB$_6$ reference powder sample (NIST SRM 660a).

All powder diffraction data (neutron and synchrotron X-ray) were analyzed using the Rietveld method, as implemented in the FullProf Suite package~\cite{FullProf_93}. The pictures of the crystal and magnetic structures were done using the VESTA software~\cite{Vesta}.

\section{State-of-the-art magnetic frustration mechanism}

In insulating oxides that are not geometrically frustrated - such as YBaCuFeO$_5$ - spiral magnetic order usually arises from the competition between nearest-neighbour $J$ and next-nearest-neighbour superexchange interactions $J'$~\cite{Coey,Blundell}. Since the setup of the spiral order is controlled by the energy scale of the weakest magnetic couplings $J'$ (typically much smaller than $J$ in insulators), this leads naturally to low $T_{spiral}$ values. In this context both the existence of spiral magnetic order in YBaCuFeO$_5$ far beyond 300~K and the positive impact of the Cu/Fe disorder in the spiral stability are unexpected observations, and suggest that a different, non-conventional mechanism should be at the origin of the magnetic frustration in this material.

A rationalization of these observations was recently proposed in two theoretical papers by Scaramucci and co workers~\cite{PRX.8,scaramucci2016spiral}. Using Density Functional Theory (DFT) calculations these authors first demonstrated the strong preference of the bipyramidal units for hosting Cu-Fe pairs, which to be consistent with the reported variability of the Cu and Fe occupations, should be disordered in the structure ~\cite{Morin1}. They also employed this technique to calculate the nearest-neighbor (NN) magnetic exchanges $J_1$, $J_2$ and $J_{ab}$ for different Cu/Fe distributions (Fig.~\ref{Structure}), which revealed that the only ferromagnetic (FM) NN exchange is the Cu-Fe coupling inside the bipyramids ($J_2$) ~\cite{Morin1}. Interestingly, all the remaining NN exchanges ($J_1$ and $J_{ab}$) are antiferromagnetic (AFM), and their sign is independent of the Cu/Fe distribution. The magnetic structure expected from the calculated exchanges, described by the propagation vector \textbf{k$_{c1}$}~=~$(\frac{1}{2}, \frac{1}{2}, \frac{1}{2})$ and shown in Fig.~\ref{Structure}a is indeed experimentally observed, but only at high temperatures ($T_{spiral}$~$<$~$T$~$<$~$T_{coll1}$).

To explain the appearance of frustration at lower temperatures ($T$~$<$~$T_{spiral}$) Scaramucci and co-workers investigated the impact of introducing a few, randomly distributed Fe-Fe `defects' in the bipyramids normally occupied by Cu-Fe pairs, accompanied by the same amount of Cu-Cu defects to keep the electric neutrality. According to their DFT calculations, such defects are energetically very expensive. However, the presence of small amounts cannot be disregarded in real samples, in particular in those with large amounts of Cu/Fe disorder. Since the $J_{Cu-Cu}$ exchange is extremely weak, introducing a few Cu-Cu defects does not perturb appreciably the underlying collinear order. In contrast, the $J_{Fe-Fe}$ coupling between the two Fe$^{3+}$  moments in a Fe-Fe defect is AFM and much stronger that the weakly FM $J_{Cu-Fe}$ coupling. A tiny amount of dilute, randomly distributed defects (2$\%$~to~6$\%$ of the bipyramids occupied by Fe-Fe pairs) constitutes thus a strong perturbation of the collinear order in the vicinity of these defects. As shown in refs.~\cite{PRX.8} and~\cite{scaramucci2016spiral}, such a perturbation does not turn the collinear order into a spin glass. Instead, it stabilizes a spiral where the collinearity is lost exclusively within the bipyramidal units, and where both, $T_{spiral}$ and the parameter $q$ related with the spiral periodicity are proportional to the Fe-Fe defect concentration.

\begin{figure}[H]
	\centering
	\includegraphics[width=0.45\textwidth]{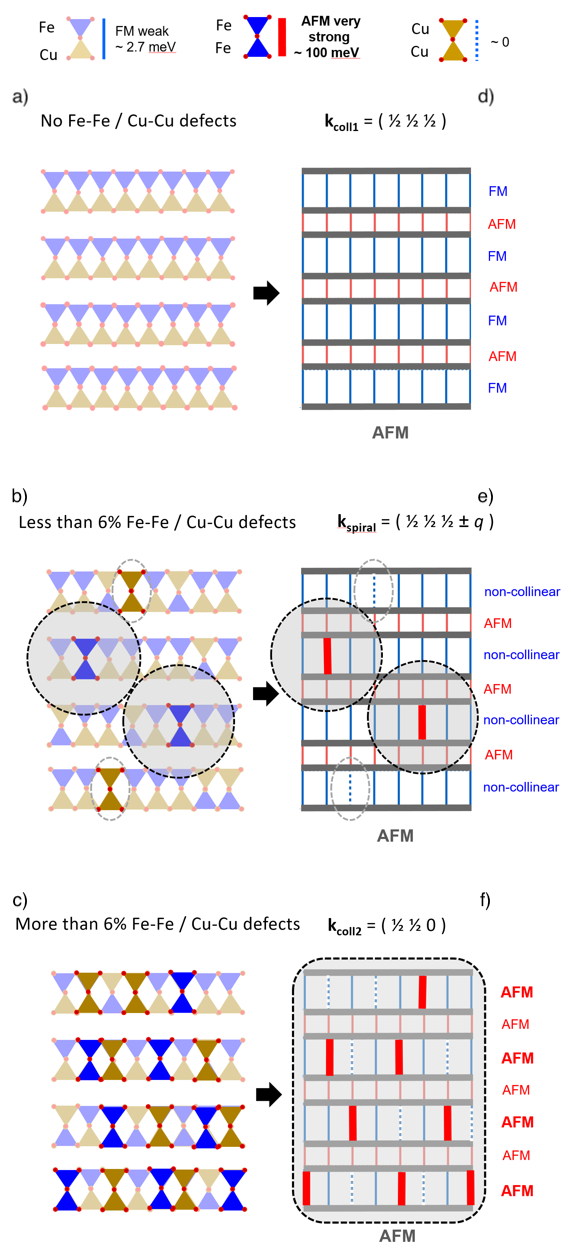}
	\caption{Schematic representation of the random exchanges frustration model illustrating the impact of increasing amounts of Cu/Fe disorder on the YBaCuFeO$_5$ magnetic order. a) Perfect Cu/Fe order. b) Medium Cu/Fe disorder with a small amount of Fe-Fe and Cu-Cu defects. c) Sizable Cu/Fe disorder with a large amount of Fe-Fe and Cu-Cu defects. d-f) Representation of a), b) and c) in terms of magnetic bonds together with the resulting \textbf{k}-vectors. Thinner bonds represent weak AFM or FM couplings, while thicker bonds indicate strong AFM couplings. Grey shaded areas indicate the regions where the Cu-Fe spins lose their FM alignment. In b) and e) the cross-talking of these regions gives rise to a spiral if the in-plane separation of the Fe-Fe impurities is smaller than xmin. In c) and f) the large number of strongly AFM Fe-Fe defects suppresses the magnetic frustration and the magnetic order becomes purely AFM. Note that the same number of weakly interacting Cu-Cu `defects' is necessary to preserve the material’s stoichiometry and charge neutrality. The J values are those of Scaramucci et al.~\cite{PRX.8,scaramucci2016spiral}.}
	\label{Model}
\end{figure}

Scaramucci and co-workers do not address in detail what will happen when the number of Fe-Fe defects becomes too important, and/or when their strength is enhanced beyond a certain limit. However, previous experimental works~\cite{TianScienceM,Romaguera_2022} point towards an abrupt suppression of the spiral order, a possibility also considered as likely within the framework of the random magnetic exchanges model. The spiral suppression would thus happen when the weak FM  Cu-Fe bonds will no longer be able to compete with the strong AFM Fe-Fe defects, leading to the suppression of the magnetic frustration. The resulting magnetic structure $coll2$, shown in Fig.~\ref{Structure}c, will be described by the magnetic propagation vector \textbf{k}$_{c2}$~=~($\frac{1}{2}$,~$\frac{1}{2}$,~0), and at variance with the high temperature $coll1$ phase, will only contain AFM nearest-neighbor couplings (Fig.~\ref{Model}c).

To date, most of the predictions of the random exchange model show an excellent agreement with the available experimental data. This provides strong support to one of its main premises - the appearance of small but growing amounts of Fe-Fe defects by increasing the degree of average Cu/Fe disorder, very difficult to probe experimentally. Moreover, it rationalizes the positive correlation between $T_{spiral}$ and the $d_{1}/d_{2}$ ratio observed experimentally ~\cite{Morin2,TianScienceM,Romaguera_2022}, where $d_{2}$ is the bipyramid size and $d_1$ is the separation between the bipyramid slabs (Fig.~\ref{Structure}a). As mentioned in the introduction, creating large amounts of Fe-Fe defects is experimentally challenging. This rule provides thus the possibility of exploiting the inverse correlation between exchange constants and interatomic distances for enhancing the impact of a $J_{Fe-Fe}$ defect (and hence $T_{spiral}$) by increasing the $d_{1}/d_{2}$ ratio.

\section{Evolution of the crystal structure}

Two space groups (SG) have been employed in the past to describe the crystal structure: $P4/mmm$ and $P4mm$, both with one unit formula per unit cell ($Z$~=~1) ~\cite{Atanassova_1993,Caignaert_1995}. $P4/mmm$ is centrosymmetric, has a mirror plane~$\perp$~\textbf{c} that contains the Y$^{3+}$ cations, and its use implies that Cu and Fe are equally distributed within the pyramids  ($n_{Cu}$~=~$n_{Fe}$~=~50$\%$). In non-centrosymmetric $P4mm$ the mirror plane is lost and the Cu/Fe occupations can be refined, allowing to accommodate asymmetric Cu/Fe distributions ($n_{Cu}$~$\neq$~$n_{Fe}$).  Although it was a long-standing polemic in the past around the actual SG, the most recent studies indicate that the Cu/Fe distribution is in general asymmetric, signalling $P4mm$ as the best suited SG for describing the crystal structure. ~\cite{Morin2,Morin1,TianScienceM,Zhang_2021,Romaguera_2022,Zhang_2022, Marelli_2024} Note that the $P4/mmm$ constrains can also be reproduced with $P4mm$ as a particular case. Disentangling both scenarios requires nevertheless high resolution powder diffraction data, either neutron or synchrotron X-ray, the Cu/Fe contrast being quite similar for both types of radiation.

Here, we employed $P4mm$ and the MSBL synchrotron XRPD data recorded at RT to access the Cu/Fe distributions of our eleven YBa$_{1-x}$Sr$_x$CuFeO$_5$ samples. The Cu/Fe disorder was described by splitting the atomic position (1/2, 1/2, $z$) inside the two pyramids of the unit cell (Fig.~\ref{Structure} and Supplemental Material). The obtained results, that will be presented in the next sections, confirm the presence of asymmetric Cu/Fe distributions in nearly all samples ($n_{Cu}$~$\neq$~$n_{Fe}$). They also reveal important degrees of Cu/Fe intermixing,  far from a full Cu/Fe order scenario with $n_{Cu}$~=~100$\%$ and $n_{Fe}$~=~0 (or $n_{Cu}$~=~0 and $n_{Fe}$~=~100$\%$). It is worth mentioning that the Rietveld fits are not sensitive to occupational correlations. The Cu/Fe disorder of our samples is thus assumed to be random, and the refined $n_{Cu}$ and $n_{Fe}$ values represent average values over the full sample volume.

Isotropic Debye-Waller factors were used for all atomic positions, a choice motivated by the presence of one A-site (Ba/Sr at $1a$) and two B-sites (Cu/Fe, both at $2h$) with chemical disorder. The $z$ positions of the basal oxygen sites O2 and O2' were fitted separately, but their mean-square displacements (MSD) were restricted to be identical. For Cu$^{2+}$ and Fe$^{3+}$, the MSD's were also constrained to have the same value. In contrast, the $z$ coordinates of the positions that these cations can occupy in the two pyramids of the unit cell were constrained as $z$(Cu1)~+~$z$(Cu2)~=~1 and $z'$(Fe1)~+~$z'$(Fe2)~=~1, respectively.

\begin{figure}[H]
	\centering
	\includegraphics[width=0.43\textwidth]{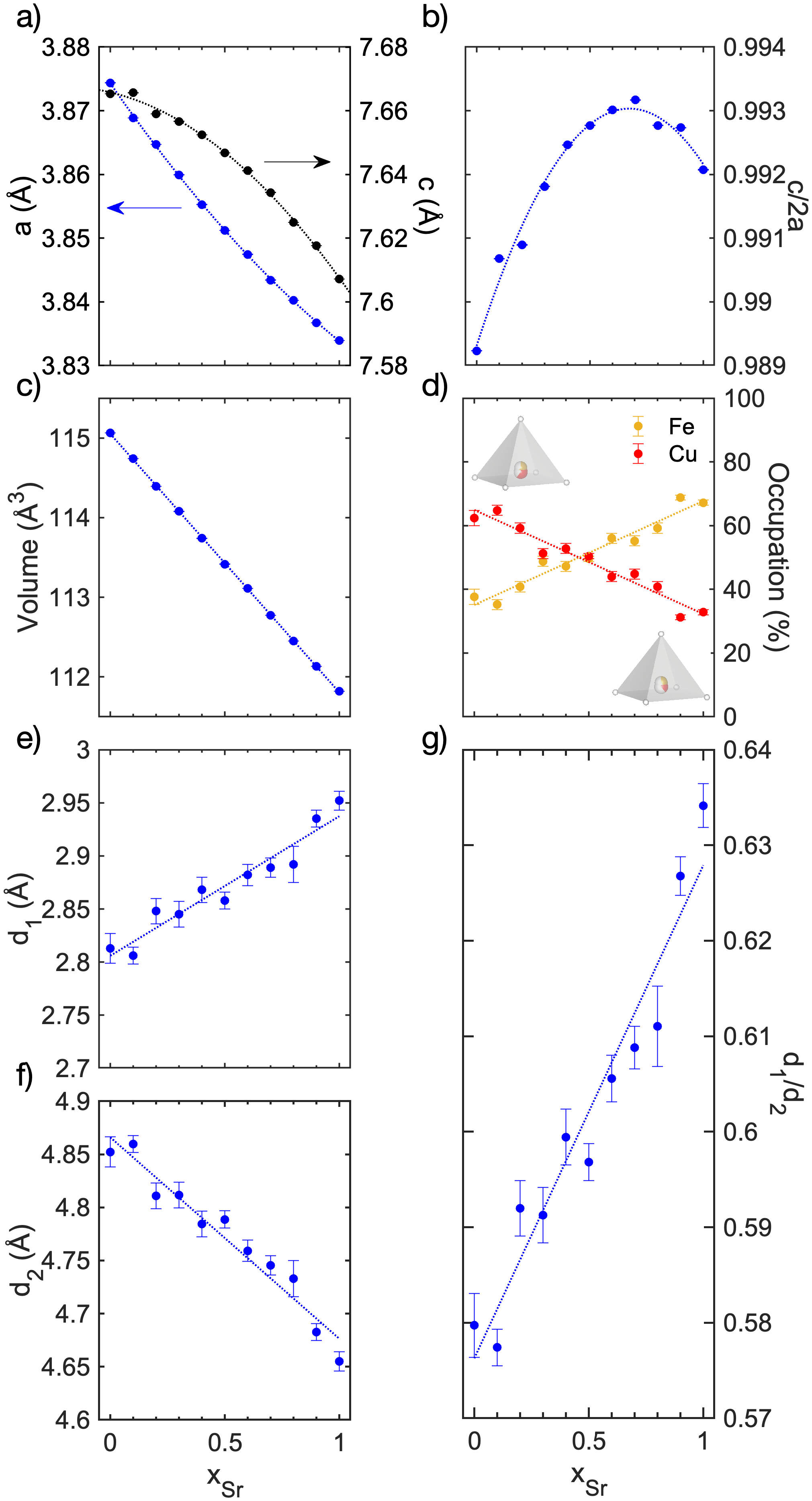}
	\caption{Evolution of the crystal structure with the Sr content. a) displays the evolution of the \textbf{a} and \textbf{c} lattice parameters, while b) and c) show the $c/2a$ ratio and unit cell volume, respectively. d) Cu/Fe occupation of the split B-sites within the upper pyramid of the unit cell together with representations of the Cu/Fe occupation for the $x$~=~0 (top left) and $x$~=~1 (bottom right) compositions. Impact of the strontium substitution on e) d$_{1}$, f) d$_{2}$ as well as their ratio g). The dotted lines are guides to the eye.}
	\label{cell+Occ}
\end{figure}

\subsection{Lattice parameters}

Figs.~\ref{cell+Occ}a and Supplemental Tables 1-3 show the modification of the lattice parameters \textbf{a} and \textbf{c} upon replacement of Ba with Sr (Supplemental Figs. 1-11, and Supplemental Tables I-III). As expected, both decrease with the Sr content $x$. However, their evolution along the series differs from the linear increase expected from Vegard’s law~\cite{Vegard} observed in other structurally-related perovskites \cite{Guillaume_94}. This deviation is better appreciated in Fig.~\ref{cell+Occ}b, showing the changes in the tetragonal distortion $c/2a$ of the pseudocubic unit cell with $x$ where $c/2a$~=~1 corresponds to a layered perovskite containing two perfectly cubic perovskite units. We note that $c/2a$ takes values smaller than one for all samples, indicating a tensile distortion that decreases with $x$ for low Sr contents . Interestingly, this tendency is reversed beyond $x$~$\simeq$~0.7, where $c/2a$ decreases with $x$. The behaviour contrasts with that of the unit cell volume $V$ (Fig.~\ref{cell+Occ}c), which contracts in a perfectly linear way with the Sr content.

\subsection{Ba/Sr and Cu/Fe distribution}

To rationalize these observations we examined the evolution of several structural parameters along the series, and compared them with the behaviour of $c/2a$. We inspected first the refined Sr and Ba fractions, and found them to agree very well with their nominal values for all samples (Supplemental Tables~I-III). Moreover, we could not find any evidence for long-range Ba/Sr order (Supplemental Fig.~1-12), even for the samples with rational Sr contents (such as $x$ = 1/2), a priory more prone to Ba/Sr ordered patterns. Also, we did not find any signature of (Ba/Sr)/Y intermixing, observed in layered perovskites when the ionic radii of the A-cations that occupy the two ordered A-sites becomes increasingly similar \cite{Marelli_2024}. An homogeneous Ba/Sr distribution (that we assume to be random in our Rietveld fits) provides thus a good  description of the experimental data, even if the presence of short-range correlations between the Ba and Sr site occupations cannot be completely disregarded.

Next, we examined the evolution of the Cu/Fe occupations $n_{Cu}$ and $n_{Fe}$ of the square-pyramidal sites, shown in Fig.~\ref{cell+Occ}d for the upper pyramid of the crystal unit cell (for the lower pyramid $n_{Cu}$ and $n_{Fe}$ are just reversed). As mentioned in the introduction, we aimed at similar Cu/Fe distributions for all samples. However, this goal was only reached for the samples with Sr contents close to 0.3~$\leq~x~\leq$~0.7, where  $n_{Cu}$~$\simeq$~$n_{Fe}$~$\simeq$~50$\%$ (Fig.~\ref{cell+Occ}d). For the remaining samples we observe a tendency towards a more ordered Cu/Fe distribution, with larger differences between the Cu and Fe occupations of a given pyramid by approaching the two end members of the series ($x$~=~0 and 1). Our analysis also reveals an unexpected trend, namely, the reversal of the Cu/Fe majority occupation by moving along the series. As shown in Fig.~\ref{cell+Occ}d, the probabilities of finding Cu and Fe in the upper pyramid of YBaCuFeO$_5$ are $\simeq$ 60$\%$ and 40$\%$, respectively (and vice versa in the lower pyramid). Interestingly, these probabilities are $\simeq$~31$\%$ and 69$\%$ for YSrCuFeO$_5$. We note also that, for this material, the higher average Fe content of the upper pyramid is consistent with its shorter height $H$~=~2.320~\AA, to be compared with $H$~=~2.335~\AA~for lower pyramid hosting a higher average Cu content (Supplemental Table I-III).

Another interesting observation is that the reversal of the Cu/Fe majority occupation starts very close to $x$~=~0.7, the Sr concentration where the $c/2a$ evolution with $x$ changes tendency. The reasons behind the Cu/Fe occupation reversal with growing Sr contents are presently unclear, but may be related by the different stability of the FeO$_5$-CuO$_5$ bipyramidal dimers in presence of Ba$^{2+}$ and Sr$^{2+}$. These A-cations are indeed located between the FeO$_5$ and CuO$_5$ pyramids, and have an impact on its size. DFT calculations for the two end members of the series, out of the scope of this work,  will be necessary to settle whether this is or not the case.

\subsection{Interatomic distances $d_1$ and $d_2$}

As mentioned in the previous section, the bipyramid size $d_1$ and the separation between the bipyramid layers $d_2$ control the strength of the exchange couplings along the \textbf{c} axis. Moreover, their ratio $d_1$/$d_2$ has been found to correlate positively with $T_{spiral}$ and inversely with $T_{coll1}$ \cite{Morin2}. Figs.~\ref{cell+Occ}e-f show the evolution of $d_1$ and $d_2$ for the full  YBa$_{1-x}$Sr$_x$CuFeO$_5$ series as a function of Sr content. As expected, $d_2$ decreases linearly with $x$. In contrast, $d_1$ increases, albeit to a slightly slower rate. This leads to an increase of $d_1$/$d_2$ upon replacing Ba with Sr (Fig.~\ref{cell+Occ}g), that we expect to result in the simultaneous increase of the spiral ordering temperature. We also note that, contrarily to the $c/2a$ ratio and the Cu/Fe occupation, the evolution of $d_1$/$d_2$ with $x$ is approximately linear and does not display any noticeable anomaly.

\section{Evolution of the magnetic order}

\subsection{Magnetic ordering temperatures}

\begin{figure*}[!hbt]
	\centering
    \includegraphics[keepaspectratio=true,width=158mm]{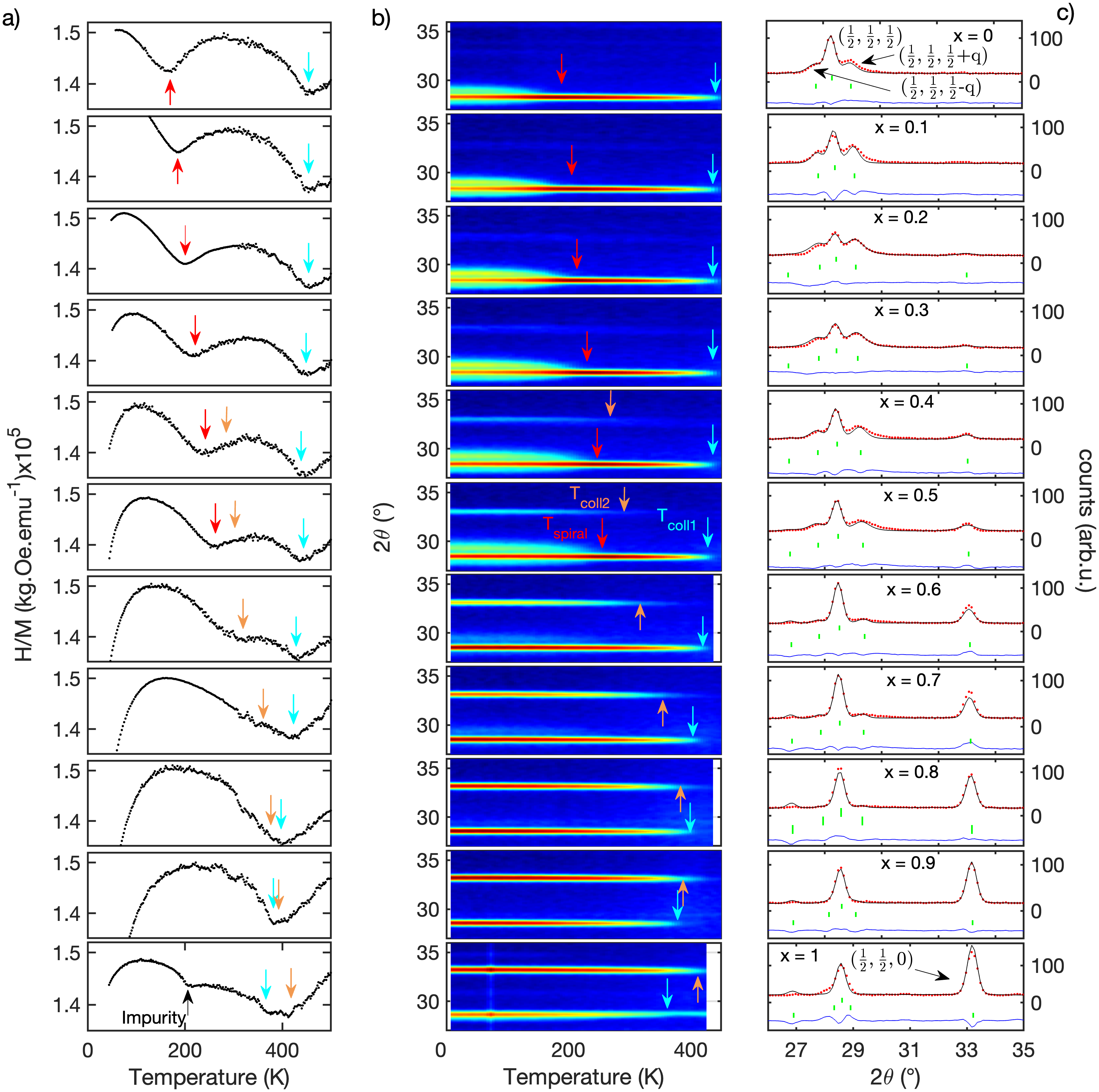}
	\caption{a) Temperature dependence of the inverse magnetic susceptibility. b) Contour plots of a selected region in the neutron powder diffraction patterns showing the thermal evolution of the ($\frac{1}{2}$, $\frac{1}{2}$, $\frac{1}{2}$), ($\frac{1}{2}$, $\frac{1}{2}$, $\frac{1}{2}$$\pm q$) and ($\frac{1}{2}$, $\frac{1}{2}$, 0) magnetic reflections related to the coll1, spiral and coll2 phases, respectively. c) Detail of the diffraction patterns measured at 2~K. Blue, red and orange vertical arrows indicate the $T_{coll1}$, $T_{spiral}$ and $T_{coll2}$ temperatures defined in the text.}
	\label{BigFig}
\end{figure*}

Fig.~\ref{BigFig}a shows the inverse magnetic susceptibility of the eleven YBa$_{1-x}$Sr$_x$CuFeO$_5$ samples. The measurements revealed the presence of (at least) two transitions in all samples at temperatures that we define here as those corresponding to minima in the $\chi^{-1}(T)$ curves. The presence of magnetic transitions was subsequently confirmed by the thermal evolution of the NPD data (Fig.~\ref{BigFig}b), which also allowed a precise assignment of the anomalies observed in the $\chi^{-1}(T)$ curves.

\begin{figure}[H]
	\centering
	\includegraphics[width=0.51\textwidth]{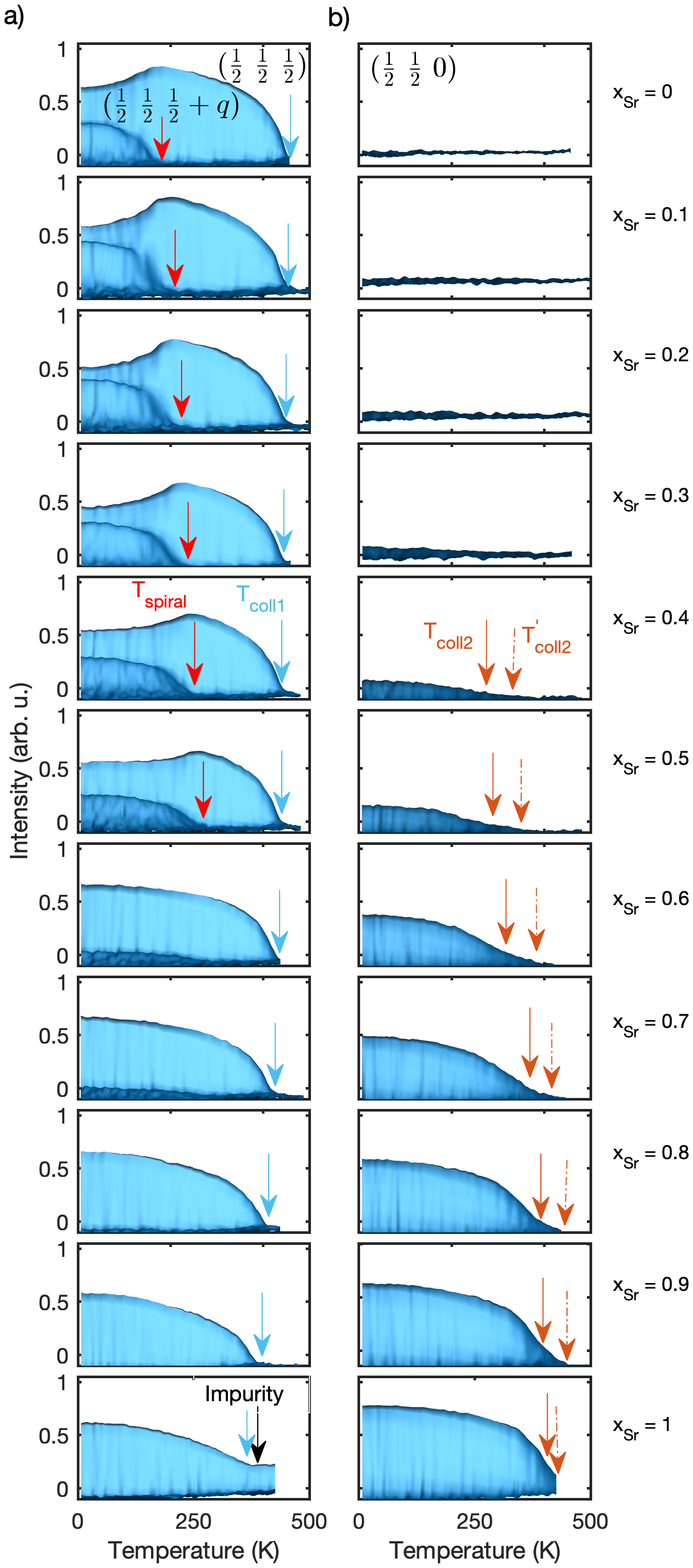}
	\caption{Thermal evolution of the magnetic reflections of the coll1, spiral and coll2 phases illustrating their different temperature dependencies in the vicinity of $x$~=~0.5. a)  ($\frac{1}{2}$, $\frac{1}{2}$, $\frac{1}{2}$) and ($\frac{1}{2}$,  $\frac{1}{2}$, $\frac{1}{2}$+$q$). b) ($\frac{1}{2}$, $\frac{1}{2}$, 0). Blue, red and orange vertical arrows indicate the $T_{coll1}$, $T_{spiral}$, $T_{coll2}$ and $T^{'}_{coll2}$ temperatures defined in the text. For a given composition, the magnetic intensities in a) and b) are shown on the same vertical scale.}
	\label{3Dint}
\end{figure}

\begin{figure}[H]
	\centering
	\includegraphics[width=0.49\textwidth]{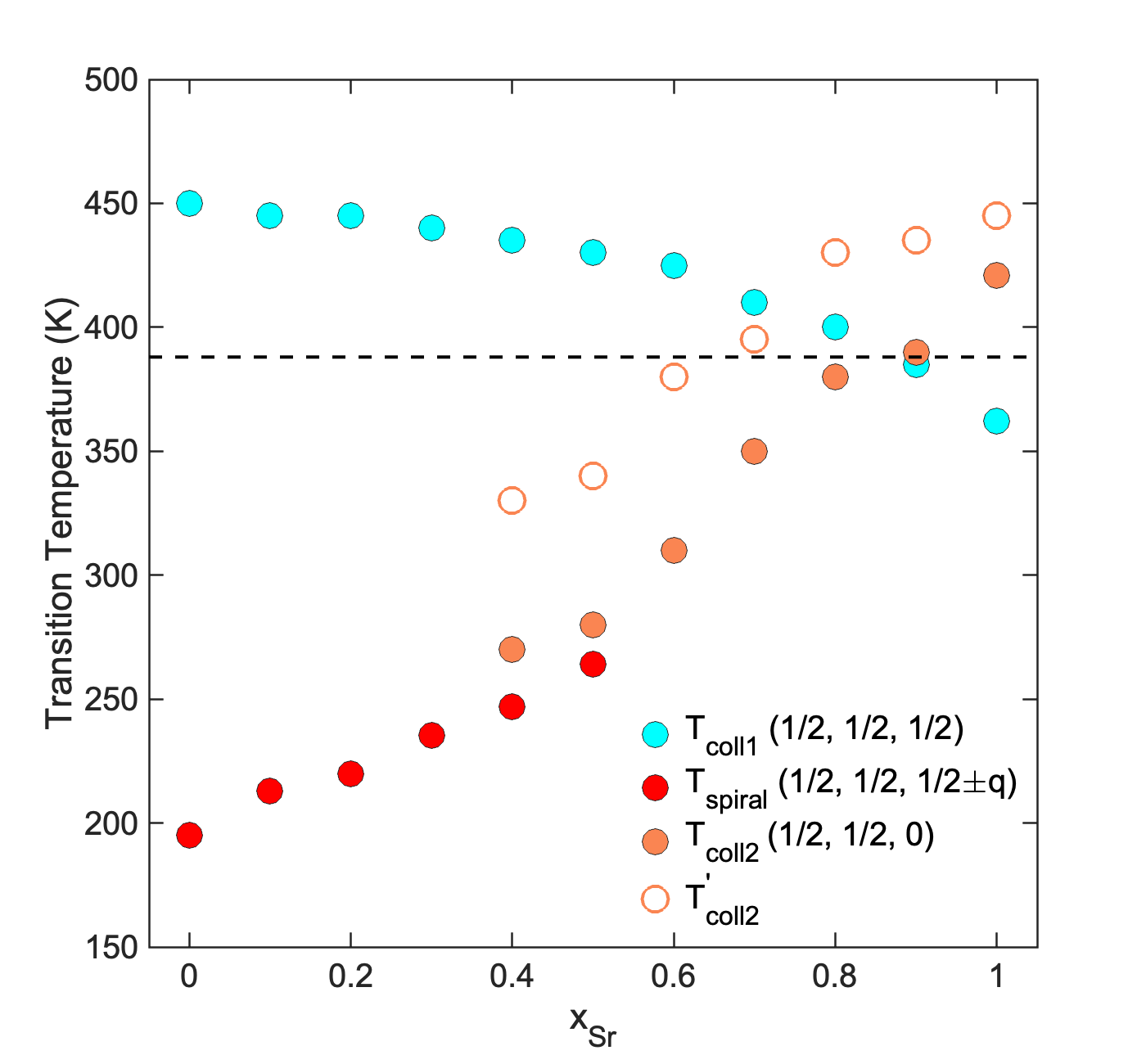}
	\caption{Evolution of the magnetic ordering temperatures $T_{coll1}$, $T_{spiral}$ and $T_{coll2}$ and the setup of the $cool2$ magnetic correlations $T'_{coll2}$ as function of the strontium content. The horizontal dashed line indicates the temperature of the paramagnetic-$coll1$-spiral triple point reported in previous studies.}
	\label{TransTemps}
\end{figure}

For the samples with 0~$\leq~x~\leq$~0.5, a comparison between the two sets of data identifies the high temperature anomalies (blue arrows in Figs.~\ref{BigFig}a-b) as the ordering temperatures of the $coll1$ phase described by the propagation vector \textbf{k$_{c1}$}~=~$(\frac{1}{2}, \frac{1}{2}, \frac{1}{2})$. It also assigns the low temperature anomalies (red arrows in Figs.~\ref{BigFig}a-b) to the ordering temperatures of the spiral phase, which coincide with a maximum in the ($\frac{1}{2}$,$\frac{1}{2}$,$\frac{1}{2}$) reflection and with the simultaneous appearance of the incommensurate satellites described by the propagation vector \textbf{k$_{s}$}~=~$(\frac{1}{2},  \frac{1}{2},  \frac{1}{2}\pm\textit{q})$ (Fig.~\ref{3Dint}a). The evolution of the $coll1$ and spiral ordering temperatures, shown in Fig.~\ref{TransTemps}, reveals a decrease of $T_{coll1}$ with $x$, whereas $T_{spiral}$ displays the opposite trend, in excellent agreement with the behaviour inferred  from the changes in $d_1$, $d_2$ and the $d_1$/$d_2$ ratio upon replacement of Ba with Sr (Figs.~\ref{cell+Occ}e-g).

We note that contrarily to the behaviour observed in the samples with the same compositions prepared at higher temperatures~\cite{TianScienceM}, the intensity of the reflections associated to the $coll1$ phase do not vanish completely below $T_{spiral}$ (Fig.~\ref{BigFig}c). This observation, reported also in early works \cite{Caignaert_1995}, can be rationalized within the framework of the random exchanges model as arising from an insufficient amount of Fe-Fe defects. Due to the low annealing temperatures employed in this study, these defects are presumably absent or too diluted in certain sample regions, leading to the preservation of the high temperature collinear phase in these areas down to the lowest temperatures.

For the samples with 0.6~$\leq~x~\leq$~1, the temperature dependence of the NPD patterns is substantially different.  Starting with the $coll1$ phase, we note that its associated reflection ($\frac{1}{2}$, $\frac{1}{2}$, $\frac{1}{2}$) is present for all compositions. However, both the setup temperature $T_{coll1}$ and the intensity of that reflection pursue the trend observed in the samples with lower Sr contents, decreasing continuously with $x$ (Figs.\ref{BigFig}b, ~\ref{3Dint}a and ~\ref{TransTemps}). 

The behaviour of the reflections associated to the spiral phase follow a completely different pattern. As shown in Fig.~\ref{BigFig}b, the incommensurate satellites become extremely broad, leading to the suppression of the maximum in the ($\frac{1}{2}$, $\frac{1}{2}$, $\frac{1}{2}$) reflection's temperature dependence, that was signaling the setup of the long-range spiral order in the Sr-poor samples (Fig.~\ref{3Dint}a). The low temperature minimum in the $\chi^{-1}(T)$ curves becomes also shallow, consistent with the evolution of the incommensurate satellites. These observations strongly suggest the absence of long-range spiral order for the samples $x~\geq$~0.6. Beyond this Sr content the signature of the spiral phase is still visible in the NPD patters for Sr contents up to $x$~=~0.8 (Fig.~\ref{3Dint}b), but it is extremely weak and broad, suggesting that the spiral magnetic order survives in the form of short-range magnetic correlations. For $x$~=~0.6, 0.7 and 0.8 the statistics of our data do not allow a precise determination of the temperature where the magnetic correlations disappear.  Interestingly, for $x$~=~0.8 (the last Sr content where the signature of these correlations is still visible) the ordering temperature of the $coll1$ phase is close to 400~K. Since this temperature coincides with the triple point reported in previous studies (Fig.~\ref{TransTemps}, \cite{TianScienceM, Romaguera_2022}), our observations suggest that the magnetic spiral cannot survive beyond 400~K, neither as long-range ordered phase, nor in the form of short-range correlations.

The abrupt disappearance of the long-range spiral order for Sr contents larger than $x$~=~0.5 is accompanied by an equally abrupt stabilization of the $coll2$ phase. This is illustrated by the growing intensity of the ($\frac{1}{2}$, $\frac{1}{2}$, 0) reflection, present but barely visible for smaller Sr concentrations, and by the increase of its onset temperature (Fig.~\ref{3Dint}b). This onset being unusually extended, we define its appearance as being between $T_{coll2}$, the temperature where the intensity of the ($\frac{1}{2}$, $\frac{1}{2}$, 0) reflection changes slope, and $T^{'}_{coll2}$, where its intensity is practically gone. The behaviour of the spiral and $coll2$ phases thus contrasts with that of the $coll1$ phase, where both, the ordering temperature $T_{coll1}$ and the intensity of the Bragg reflections evolve smoothly across $x$~=~0.5 (Fig.~\ref{TransTemps}).

These observations suggest a positive impact of the Sr content on the stability of the $coll2$ phase, which replaces the spiral for $x$~$>$~0.5 and grows slowly at expenses of a $coll1$ phase progressively destabilized upon replacing Ba with Sr. This behaviour is consistent with the continuous increase of $J_{Fe-Fe}$ anticipated from the $d_2$ contraction with $x$ (Fig.~\ref{cell+Occ}f), which within the framework of the random exchanges model is expected to suppress the magnetic frustration and stabilize the purely AFM order described by the propagation vector \textbf{k$_{c2}$} = $(\frac{1}{2}, \frac{1}{2}, 0)$.

\begin{figure}[H]
	\centering
	\includegraphics[width=0.42\textwidth]{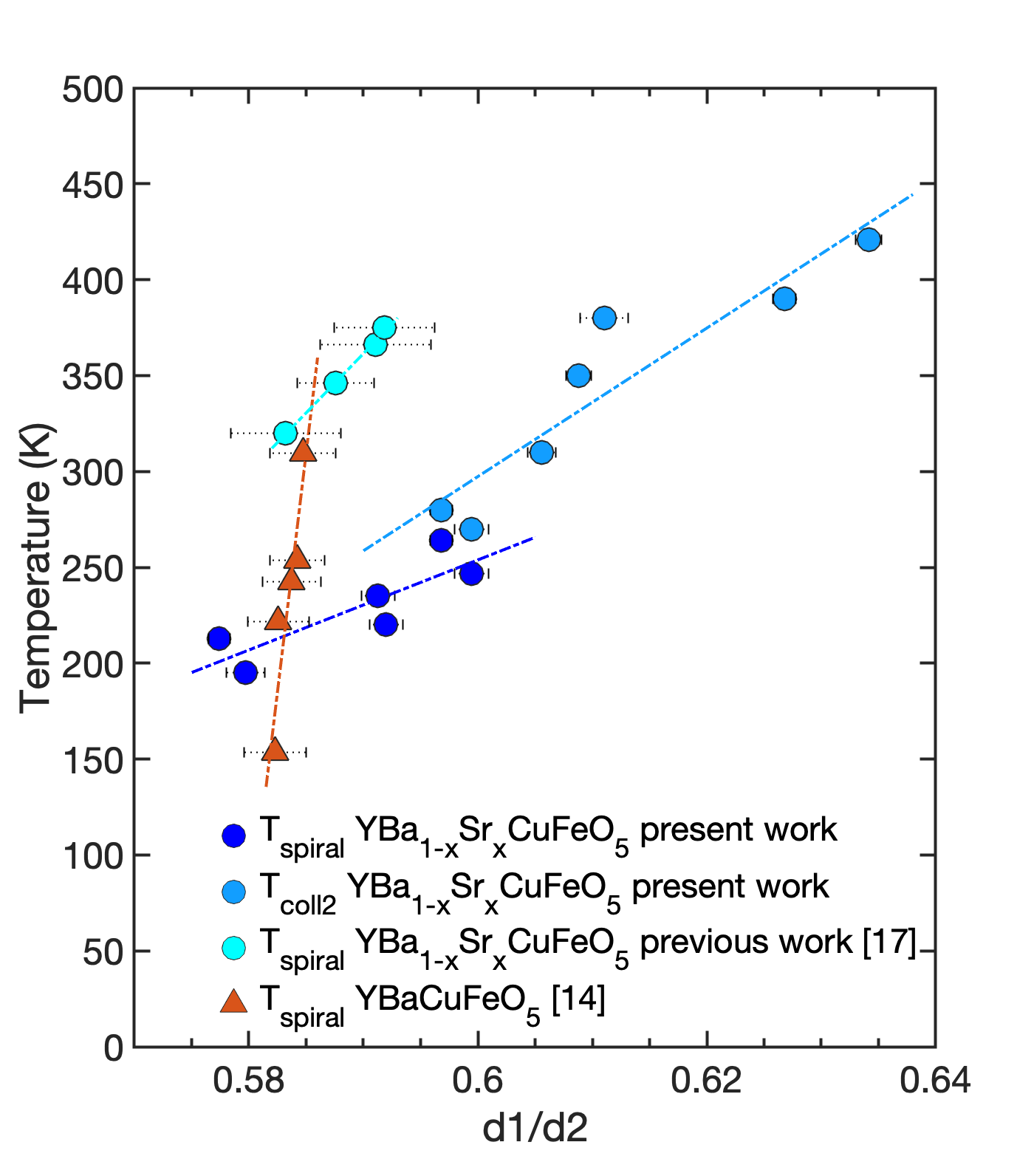}
	\caption{Evolution of the spiral transition temperature as function of the $d_{1}$/$d_{2}$  ratio. The figure includes present and previously reported data. Dark blue points represent the data reported in the work with samples from $x$~=~0 to $x$~=~0.5, while the light blue points correspond to the transition temperatures for the second collinear phase ($\frac{1}{2}$,~$\frac{1}{2}$,~0).}
	\label{TvsD1D2}
\end{figure}

We would like to comment briefly the possible reasons behind the $coll1$/$coll2$ phase coexistence observed for the samples with $x$~$>$~0.5. This phenomenon is also observed in the samples with low Sr contents, were the spiral and $coll1$ phases coexist for all the samples investigated. However, we know from our previous studies that this phase coexistence is not intrinsic and can be avoided by employing synthetic routes that improve the homogeneity of the Cu/Fe distribution (such as high annealing temperatures of soft chemistry \cite{Morin1,Morin2,TianScienceM}). Given that the region of the phase diagram beyond the triple point has not been theoretically investigated in detail, neither theoretically nor experimentally, the $coll1$/$coll2$ phase coexistence observed for $x$ $>$ 0.5 could have a different origin. In this context, DFT calculations for a given Sr content in presence of the $coll1$ and $coll2$ magnetic orders could help to establish the relative stability of the two magnetic phases. The low, identical annealing temperature employed to prepare the samples investigated in this study suggest however that the possibility of avoiding phase coexistence through a different synthesis route cannot be disregarded.  Finding whether this is or not the case will nevertheless need additional experimental and theoretical work

We focus now on the $coll2$ phase, less investigated in previous studies, where the NPD data reveal strong differences in the temperature dependence of its associated ($\frac{1}{2}$,  $\frac{1}{2}$, 0) reflection for different Sr contents. As shown in Fig.~\ref{3Dint}b, the decay of this reflection is relatively fast for $x$~=~1, but it becomes increasingly slow by decreasing the Sr content. Moreover, it displays a curvature change at temperatures $T_{coll2}$, substantially lower than the onset temperature $T'_{coll2}$ (full and dashed arrows, respectively) accompanied by extra broadening. A detailed analysis of the critical exponents and correlation lengths in the region between $T'_{coll2}$ and $T_{coll2}$, that we anticipate to be highly unconventional, is out of the scope of this work and will be addressed separately. However, the remarkable difference of the ($\frac{1}{2}$, $\frac{1}{2}$ 0) decay with that of the reflection ($\frac{1}{2}$, $\frac{1}{2}$, $\frac{1}{2}$), associated to a magnetic phase whose volume fraction does not change abruptly across $x$~=~0.5, suggests the presence of short-range magnetic correlations between  $T'_{coll2}$ and $T_{coll2}$.

Assuming that fully long-range $coll2$ magnetic order is realized only below $T_{coll2}$, we have plotted the evolution of this temperature with the Sr content together with that of $T'_{coll2}$, $T_{coll1}$ and $T_{spiral}$. The results, shown in Fig.~\ref{TransTemps}, reveal two interesting trends. Firstly, $T_{coll2}$ increases with $x$ at a rate $\partial$$T_{coll2}$/$\partial$$x$ very similar to that of $T_{spiral}$ in the $x~\leq~0.5$ region. Secondly, the $T_{coll1}(x)$  and $T_{coll2}(x)$ lines cross exactly at $T_{coll1}$~=~$T_{coll2}$~$\sim$~400~K, i.e., the temperature of the paramagnetic-$coll1$-spiral triple point reported in previous studies. These observations strongly suggest a common origin for the spiral and $coll2$ phases, a scenario also supported by the evolution of $T_{spiral}$ and $T_{coll2}$ with the $d_{1}$/$d_{2}$ ratio. As shown in Fig.~\ref{TvsD1D2}, both temperatures are linear functions  of $d_{1}$/$d_{2}$ and increase with this variable at similar rates.

\begin{figure*}[!hbt]
\includegraphics[keepaspectratio=true,width=170 mm]{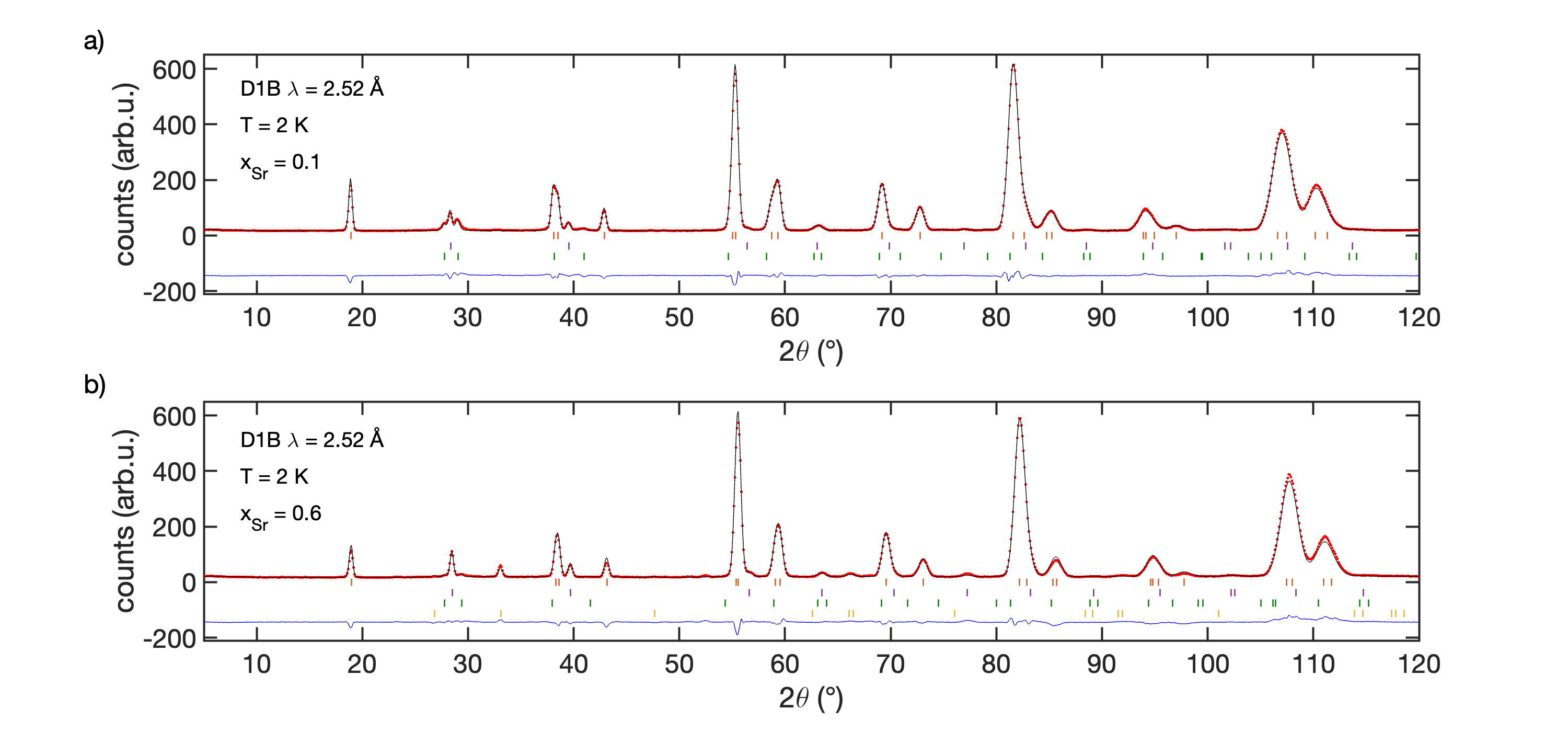}
\caption{Neutron diffraction pattern of samples x$_{Sr}$~=~0.1 (a) and x$_{Sr}$~=~0.6 (b) recorded on D1B, ILL with a 2.52~\AA \ incident wavelength.}
\label{Diffracto}
\end{figure*}

By looking a Fig.~\ref{TransTemps}, a question arises: if according to previous experimental studies the $coll2$ phase is only stable beyond a critical Fe-Fe defect concentration of $x$~$\simeq$~6$\%$, why is it observed in the samples with $x$~=~0.4 and 0.5, where the simultaneous presence of the spiral phase with $T_{spiral}$ suggest defect concentrations between 3$\%$ and 4$\%$ \cite{PRX.8}? Simultaneously, we can also wonder why for $x$~=~0.6, 0.7 and 0.8 the $coll2$ phase shows long-range order whereas the spiral phase is only present in form of short-range correlations. Again, we believe that the reason may be an inhomogeneous Cu/Fe distribution caused by the low annealing temperature, which may include Fe-rich (and Cu-rich regions) with respectively higher/lower concentrations of Fe-Fe defects. The impact of these inhomogeneities in the stability of the different magnetic phases is difficult to evaluate without a precise knowledge of the local Cu/Fe distribution, in particular close to the triple point, where the energy scale of the $coll1$, spiral and $coll2$ phases became comparable to that of the paramagnetic state.

\subsection{Magnetic structures}

After uncovering the stability of the different magnetic phases upon replacement of Ba with Sr, we address now the changes in the periodicity of the spiral phase, directly related with the incommensurate modulation parameter $q$ of the magnetic propagation vector \textbf{k$_{s}$} = $(\frac{1}{2},  \frac{1}{2},  \frac{1}{2}\pm\textit{q})$. According to the predictions of the random exchanges frustration model~\cite{PRX.8,scaramucci2016spiral}, $T_{spiral}$ should be directly proportional to the ground state value of this parameter $q_G$. Here, we extracted this information from the fits of the NPD patterns at 2~K, that we conducted for all the samples. Representative examples of such fits for the samples with long range spiral ($x$ $\leq$ 0.5) and $coll2$ order ($x$~$>$~0.5) are shown in Fig.~\ref{Diffracto}.

The collinear and spiral magnetic structures were described according to the models first reported in \cite{Morin1}. Because of the inherent limitation of the unpolarized NPD technique, the Rietveld fits could not be performed without introducing restrictions between the refined parameters. These restrictions, consistent with those employed in our previous works, and also in more recent studies, were the following: a) The spiral envelop was assumed to be circular. b) At all temperatures, the ratio between the Fe and Cu magnetic moments was restricted to be identical to their free ion, spin-only values (5:1).  c) In the temperature regions where two magnetic phases coexist, the Fe and Cu magnetic moments of the two phases were also constrained to have the same value and the same inclination with respect to the \textbf{ab} plane.

\begin{figure}[]
	\centering
	\includegraphics[width=0.48\textwidth]{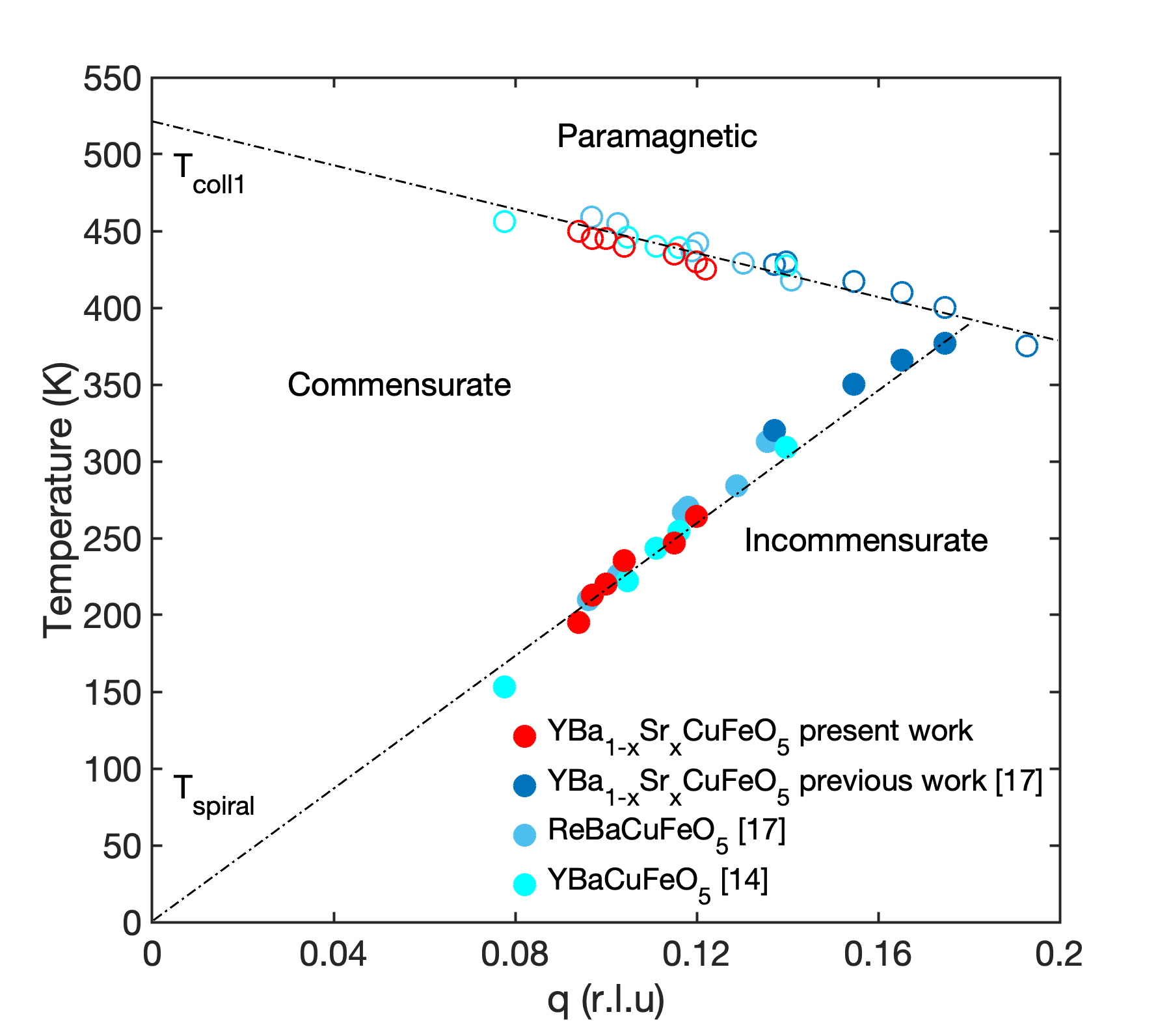}
	\caption{$T_{coll1}$ and $T_{spiral}$ magnetic transition temperatures versus $q_G$ for Cu/Fe based layered perovskites. Only samples with long-range spiral order are displayed. Red points represent the data reported in this study for the samples with 0~$\leq$~$x$~$\leq$~0.5. Filled markers stand for the commensurate to incommensurate transition whereas the empty ones depict the paramagnetic to commensurate transition.}
	\label{TvsQ}
\end{figure}

The evolution of $T_{spiral}$ as a function of the refined $q_G$ values, shown in Fig.~\ref{TvsQ}, reveals a nearly linear relationship between both variables with a slope that nicely coincides with that reported in previous studies. It is worth mentioning that the points shown in Fig.~\ref{TvsQ} include results reported by different groups. Moreover, different colors/markers correspond to different layered Cu/Fe-based perovskite families with different $d_2$ values (and hence different $J_{Fe-Fe}$ exchange constants). The observation of a common $T_{spiral}$($q_G$) law, which can be described by a straight line crossing the origin to a very good approximation, is thus a remarkable result that provides additional support to the random exchanges frustration model. Another interesting observation is that the slope of the $T_{spiral}$($q_G$) line depends of the values of the NN exchange constants, and in particular, on $J_{Fe-Fe}$. The observation of a nearly common slope with very low dispersion indicates that the variations of the exchange constants (including $J_{Fe-Fe}$) among the different materials is not very large, as inferred from some of the results presented in previous sections.

\section{Conclusions and Outlook}

The aim of this study was to investigate the region beyond the paramagnetic-collinear-spiral triple point observed in Cu/Fe-based layered perovskites upon introduction of small concentration of strongly frustrating AFM Fe-Fe defects in the bipyramidal units occupied majoritarily by weakly FM Cu-Fe pairs. Being aware of the experimental difficulties linked to the generation of large defect concentrations, we attempted an alternative route for accessing the phase diagram region beyond the triple point. We thus synthesized the YBa$_{1-x}$Sr$_{x}$CuFeO$_{5}$ solid solutions ($0~\leq~x~\leq~1$), where we progressively replaced Ba with Sr with the aim of enhancing the strength of the experimentally available Fe-Fe bonds. Our strategy was only partially successful due to difficulties associated to the synthesis procedure, which required to find a compromise between the generation of large amounts of Fe-Fe defects and the obtention of phase pure samples.

Using a combination of bulk magnetization, X-ray and neutron powder diffraction we could show that the spiral state is destabilized beyond a critical Fe-Fe defect concentration, being replaced by a non-frustrated, fully antiferromagnetic state with propagation vector \textbf{k$_{c2}$}~=~$(\frac{1}{2},~\frac{1}{2},~0)$ and ordering temperature $T_{coll2}$~$\geq$~$T_{spiral}$. Interestingly, $T_{spiral}$ and $T_{coll2}$ increase with $x$ at the same rate, suggesting a common, Fe-Fe defect-driven origin. These findings are consistent with the theoretical predictions of the random exchanges model developed by Scaramucci and co-workers, further validating this novel, disorder-based frustration mechanism as an efficient tool for the design of spiral magnets with ordering temperatures high enough to display  magnetoelectric properties beyond room temperature.

\section{Acknowledgements}
We would like to thank Mickael Morin for fruitful discussions, and Ekaterina Pomjakushina for the availability of the tubular furnace and the TG device. We aknowledge the beamtime allocation at the Swiss Light Source (Materials Science Beamline) and at the Institut Laue Langevin (powder difrractometer D1B, through the Spanish CRG access program). X. Torrelles and M. Medarde acknowledge the Spanish Science Ministry for travel funding. V. Poree acknowledges the MaMaself Erasmus Mundus program for financial support during his stay at the Paul Scherrer Institut.

\bibliography{biblio}

\end{document}